\documentstyle[prl,floats,twocolumn,aps,epsf,psfrag]{revtex}

\tighten

\renewcommand{\baselinestretch}{1}
\begin{document}
\twocolumn[\hsize\textwidth\columnwidth\hsize\csname @twocolumnfalse\endcsname
\title{Quantum no-deleting principle and some of its implications}
\author{Arun K. Pati$^{(1)}$ and Samuel L. Braunstein }
\address{Quantum Optics and Information Group,
School of Informatics, Dean Street, University of Wales, Bangor LL 57 1UT, U.K.}
\address{$^{(1)}$ Theoretical Physics Division, 5th Floor, Central Complex,
BARC, Mumbai-400 085, India}

\date{\today}
\maketitle
\def\ra{\rangle}
\def\la{\langle}
\def\ver{\arrowvert}
\begin{abstract}
Unmeasureability of a quantum state has important consequences in practical
implementation of quantum computers. Like copying, deleting of an unknown
state from among several copies is prohibited. This is called no-deletion
prinicple. Here, we present a no deleting
principle for qudits. We obtain a bound on $N$-to-$M$ deleting and show that
the quality of deletion drops exponentially with the number of copies to be
deleted. In addition, we investigate conditional, state-dependent and approximate
quantum deleting of unknown states. We prove that unitarity does not allow
us to delete copies from an alphabet of two non-orthogonal states exactly.
Further, we show that no-deleting principle is consistent with no-signalling.
\end{abstract}

\vskip .5cm



\par

Linearity of quantum theory unveils that we cannot duplicate
\cite{wz,dd} an unknown quantum state accurately.
Unitarity of quantum evolution shows that non-orthogonal quantum states
cannot be perfectly copied \cite{hy}. However, they can be
can be copied exactly by a unitary and measurement process \cite{dg,dg1}.
The possibility of copying an unknown
state approximately by deterministic \cite{bh,bbhb,bh1,nc,gm,bem} cloning
machines was also proposed.
Further, non-orthogonal states can evolve
into a linear superposition of multiple copies by a novel cloning
machine \cite{akp}.

In the quantum information and computation era it is important to know
what we can do with the vast amount of information contained in an unknown
state and what we cannot. For example, given several (a finite number of)
copies of an unknown state we can partly estimate it \cite{mp,dbe},
we can swap it and we can teleport it \cite{betal}.
But can we delete a copy of an {\em unknown state} from a collection of
several copies?

The quantum deletion we study here is not same as the erasure.
When we wish to get rid of the last bit (either classical or quantum) of
information it is called primitive erasure. This can
be achieved by spending certain amount of energy (thermodynamically irreversible),
known as Landauers's erasure principle \cite{ls,rl}. We will consistently
use the term  ``deleting'' to refer to a uncopying-type of operation as
opposed
to primitive erasure. We \cite{pb} have
recently shown that unlike classical information
in quantum theory the {\em perfect deletion} of an unknown qubit from a
collection of two or more qubits is an  impossible operation. The basic
linear structure of quantum theory puts severe limitations on the complete
deleting of the quantum information of an unknown state \cite{whz}.
In this letter we generalise our no-deleting principle to qudits
(qudit is a $d$-dimensional quantum system).
We obtain a bound on the maximum limit of $N$-to-$M$ quantum deletion.
We study so-called conditional quantum deleting and state-dependent,
approximate deleting. We prove a no-deletion theorem for non-orthogonal states
using unitarity. Interestingly, we show that the quantum no-deleting
principle is consistent with no-signalling.

{\it Quantum deleting of qudits}:
Consider several copies (say $N$) of an unknown quantum state $\arrowvert \Psi \rangle$
each in a $d$-dimensional Hilbert space ${\cal H} = {\cal C}^{d}$.
Our $N$ copies $\arrowvert\Psi \rangle^{\otimes N}$
live in a smaller dimensional subspace, which is the symmetric subspace of
${\cal H}^{\otimes N}$. It contains states that are symmetric under
interchange of any pair of qudits. The aim of the quantum deleting machine
is to delete one or more number of qudits from a collection of two or more
qudits {\em all in the same state}. In a sense, we intend to construct a
machine which appears to perform the
`reverse' of cloning operation (but as we will see later, strictly it is not
so). In general the quantum deleting operation is defined for $N$ unknown
states $\arrowvert\Psi \rangle^{\otimes N}$ such that the linear operator acts on the
combined Hilbert space and deletes  $(N-M)$ copies and keeps
$M$ copies intact. It is defined by

\begin{eqnarray}
\arrowvert \Psi \rangle^{\otimes N} \arrowvert A \rangle \rightarrow \arrowvert\Psi \rangle^{\otimes M} \arrowvert\Sigma \rangle^{\otimes (N-M)} \arrowvert A_{\Psi} \rangle,
\end{eqnarray}
where $\arrowvert\Sigma \rangle$ is the blank state of a qudit,
$ \arrowvert A \rangle$ is the initial
and $\arrowvert A_{\Psi} \rangle$ is the final state of the ancilla which may
in general depend on $\arrowvert \Psi \rangle$.

For simplicity, let us consider a $2$-to-$1$ quantum deleting machine 
for qudits.
Let $\{\arrowvert\psi_i \rangle, i=1,2,\ldots d\} \in {\cal H}$ be an arbitrary
orthonormal basis of a qudit ($\psi$ is labelling the choice of basis).
For pedagogical reason consider a scenario where only Hilbert spaces
involving the initial copies will be investigated.
Thus, the deleting machine is a linear operator that acts jointly
on the copies of qudits, given by

\begin{eqnarray}
\arrowvert\psi_i \rangle_a \arrowvert\psi_i \rangle_b  \rightarrow \arrowvert\psi_i \rangle_a \arrowvert\Sigma \rangle_b.
\end{eqnarray}
Note that if the inputs are different then the deleting machine can yield
an arbitrary state such as

\begin{eqnarray}
\arrowvert\psi_i \rangle_a \arrowvert\psi_j \rangle_b  \rightarrow \arrowvert\Phi_{ij}^{\psi} \rangle_{ab} (i \not=j).
\end{eqnarray}
However, if we send two copies of an arbitrary qudit in a state 
$\arrowvert\Psi \rangle= \sum_{i=1}^{d} c_i \arrowvert\psi_i \rangle$ (with $c_i$' s being all
{\em unknown} complex numbers), then by linearity we have

\begin{eqnarray}
\arrowvert\Psi \rangle_a \arrowvert\Psi \rangle_b  \rightarrow \sum_{i=1}^{d} c_i^2 \arrowvert\psi_i \rangle_a
\arrowvert\Sigma \rangle_b + \sum_{{ij=1}_{ i \not=j} }^{d}
 c_ic_j \arrowvert\Phi_{ij}^{\psi} \rangle_{ab}.
\end{eqnarray}
But ideally we would require
$\arrowvert \Psi \rangle_a \arrowvert \Psi \rangle_b  \rightarrow \sum_{i} c_i \arrowvert\psi_i \rangle_a
\arrowvert\Sigma \rangle_b$. Since ideal output and actual output states are different
linearity does not allow us to delete an unknown copy of a quantum state
in any finite dimensional Hilbert space.

We can also prove the quantum no-deletion principle by including ancilla. However,
when we include ancilla we have to exclude swapping of an unknown copy
onto ancilla as a proper deletion. The reason for doing so is first, swapping
of an unknown copy onto ancilla is just hiding of quantum information and second,
if we allow swapping then the result
reduces to primitive erasure and the
extra copies that are available at our disposal have played no role in deleting
mechanism. Intuitively one would say that if we have a large number of copies
of an unknown quantum state we know more about the state (as only in the
limit of infinite number of copies we know the state exactly). So extra 
copies should help in deleting an unknown state. Now including ancilla let
us define the action of deleting machine on orthogonal qudits as given by

\begin{eqnarray}
\arrowvert \psi_i \rangle_a \arrowvert \psi_i \rangle_b \arrowvert A \rangle_c \rightarrow
\arrowvert \psi_i \rangle_a \arrowvert \Sigma \rangle_b \arrowvert A_{\psi_i} \rangle_c,
\end{eqnarray}
where $\arrowvert A_{\psi_i} \rangle$'s need not be orthogonal. When the
inputs are in different state then the deleting machine can yield
an arbitrary state such as

\begin{eqnarray}
\arrowvert \psi_i \rangle_a \arrowvert \psi_j \rangle_b  \arrowvert A \rangle_c \rightarrow \arrowvert \Phi_{ij}^{\psi} \rangle_{abc}.
\end{eqnarray}
It should be noted that because of (5) and (6) quantum deleting machine is
not reverse of cloning machine.
If we send two copies of an arbitrary qudit in a state such as
$\arrowvert\Psi \rangle$, then by linearity we have

\begin{eqnarray}                               
&& \arrowvert \Psi \rangle_a \arrowvert \Psi \rangle_b
\arrowvert A \rangle_c \rightarrow \sum_{i} c_i^2 \arrowvert \psi_i \rangle_a
\arrowvert \Sigma \rangle_b  \arrowvert A_{\psi_i} \rangle_c \nonumber \\
&& + \sum_{{ij}_{ i \not=j}}
 c_ic_j \arrowvert \Phi_{ij}^{\psi} \rangle_{abc},
\end{eqnarray}
which is a qudratic polynomial in $c_i$'s. 
But ideally if we would be able to delete a copy of an unknown qudit then
the above equation must reduce to $\sum_{i} c_i \arrowvert \psi_i \rangle_a
\arrowvert \Sigma \rangle_b \arrowvert A_{\Psi} \rangle_c$ for all $c_i$'s.
Note that once we exclude swapping, the ancilla state 
$\arrowvert A_{\Psi} \rangle$ is independent of the input state
$\arrowvert \Psi \rangle$.
Since we know that
$\arrowvert \Phi_{ij}^{\psi} \rangle_{abc}$ can never depend on $c_i$'s, therefore the only
solution this allows is $\arrowvert \Phi_{ij}^{\psi} \rangle_{abc} = \arrowvert \psi_i \rangle_a
\arrowvert \Sigma \rangle_b \arrowvert A_{\psi_j} \rangle_c$ and
$\arrowvert A_{\Psi} \rangle_c = \sum_i c_j \arrowvert A_{\psi_j} \rangle_c$.
Since the final actual output state has to be normalised for all $c_i$'s we
can see that all $\arrowvert A_{\psi_j} \rangle$'s have to be orthonormal.
Therefore, linearity allows only swapping of an unknown qudit onto the $d$-dimensional subspace
of the ancilla. That is the unknown quantum state is just hidden in the
deleting machine state which can of course be retrieved by a unitary
transformation. Hence, linearity does not allow us to delete an unknown state
against a copy, even in the presence of ancilla. We can only move quantum
information around but we cannot delete it completely. This is called quantum
no-deletion principle for qudits.


{\it Bounds on $N$-to-$M$ quantum deleting:} Below we restrict our discussions
for qubits but they can be generalised to qudits.
Consider a new scenario where Victor owns a ``qubit company'' and prepares
copies of a qubit. Let $N$ copies of a qubit given to us live in a symmetric
subspace of
$(N+1)$ dimensional Hilbert space of the full $2^N$-dimensional Hilbert space.
The $N$ copies of an unknown qubit
$\arrowvert \Psi \rangle=  \alpha  \arrowvert 0 \rangle+ \beta  \arrowvert 1 \rangle$
can be written as
\begin{eqnarray}
\arrowvert \Psi \rangle^{\otimes N} =  \alpha^{N}  \arrowvert 0 \rangle^{\otimes N} +
\beta^{N}  \arrowvert 1 \rangle^{\otimes N} + \sum_{k=1}^{N-1} f_k(\alpha,\beta) \arrowvert k \rangle,
\end{eqnarray}
where $\arrowvert k \rangle$'s are $(N-1)$ orthogonal bit string states
living in the symmetric subspace.
 
Our deleting machine for orthogonal qubits is defined by
$\arrowvert 0 \rangle^{\otimes N} \arrowvert A \rangle  \rightarrow 
\arrowvert 0 \rangle^{\otimes M}
\arrowvert\Sigma \rangle^{\otimes (N-M)} \arrowvert A_0 \rangle,~~
\arrowvert 1 \rangle^{\otimes N} \arrowvert A \rangle  \rightarrow
\arrowvert 1 \rangle^{\otimes M}
\arrowvert \Sigma \rangle^{\otimes (N-M)} \arrowvert A_1 \rangle,~~$ and
$\arrowvert k \rangle \arrowvert A \rangle \rightarrow
\arrowvert k' \rangle$,
where  $\arrowvert k' \rangle$ is final state of the symmetric $N$ qubits
state and the ancilla. If we send $N$ copies of an unknown qubit through our deleting machine this
will yield

\begin{eqnarray}
&& \arrowvert \Psi \rangle^{\otimes N} \arrowvert A \rangle \rightarrow
\alpha^{N}  \arrowvert0 \rangle^{\otimes M}
\arrowvert \Sigma \rangle^{\otimes (N-M)} \arrowvert A_0 \rangle + \nonumber \\
&& \beta^{N}  \arrowvert 1 \rangle^{\otimes M} \arrowvert \Sigma \rangle^{\otimes (N-M)} \arrowvert A_1 \rangle+ 
\sum_{k=1}^{N-1} f_k(\alpha,\beta) \arrowvert k' \rangle= \arrowvert \Psi_{{\rm actual}} \rangle.
\end{eqnarray}
Now, Victor wants to test how good is our quantum deleting machine by
evaluating the worst case success of performing deletion operation.
As Victor knows what the state is he can delete $(N-M)$ copies perfectly and
keep $M$ copies intact. This ideal operation is given by

\begin{eqnarray}
&& \arrowvert \Psi \rangle^{\otimes N} \arrowvert A \rangle \rightarrow
 \arrowvert \Psi \rangle^{\otimes M}
 \arrowvert \Sigma \rangle^{\otimes (N-M)} \arrowvert A_{\Psi} \rangle =
 (\alpha^{M}  \arrowvert 0 \rangle^{\otimes M}
+ \beta^{M}  \arrowvert 1 \rangle^{\otimes M} \nonumber \\
&& + \sum_{j=1}^{M-1} g_j(\alpha,\beta) \arrowvert j \rangle) \arrowvert \Sigma \rangle^{\otimes (N-M)}
\arrowvert A_{\Psi} \rangle
= \arrowvert  \Psi_{{\rm ideal}} \rangle.
\end{eqnarray}
Therefore, the error introduced by the quantum deleting machine 
can be calculated from ${\cal E} = 1 - {\cal Q} = 1- \arrowvert \langle
\Psi_{{\rm actual}} \arrowvert
\Psi_{{\rm ideal}} \rangle \arrowvert$. If the ideal and the actual states
are identical then obviously there is no error. The quantity
${\cal Q}$ called quality function is bounded by

\begin{eqnarray}
&& {\cal Q} \le  \arrowvert \alpha \arrowvert^{(N+M)} + \arrowvert \beta \arrowvert^{(N+M)} \nonumber \\
&& + [1- (\arrowvert \alpha \arrowvert^{2N} + \arrowvert \beta \arrowvert^{2N} )]^{\frac{1}{2} }
[1- (\arrowvert \alpha \arrowvert^{2M} + \arrowvert \beta \arrowvert^{2M} ) ]^{\frac{1}{2} }.
\end{eqnarray}
The rhs of (11) can be optimised and the optimal value of ${\cal Q}$ is given by

\begin{eqnarray}
{\cal Q}_{{\rm opt}} =  \frac{2}{2^{(N+M)/2} } +
\sqrt{\bigg(1- \frac{2}{2^N} \bigg) \bigg(1 - \frac{2}{2^M} \bigg) }.
\end{eqnarray}
As expected the function ${\cal Q}_{{\rm opt}} $ is one for $N=M$,
so there is no error. For $N \rightarrow 1$ deleting (meaning keeping a
single copy and deleting $(N-1)$ copies, the quality function
${\cal Q}_{{\rm opt}} = \frac{1}{2^{(N-1)/2} }$.
Interestingly, the quality goes down exponentially with number of copies we
would like to delete, hence the error increases. Therefore, it is difficult
to delete more and more number of copies. From (12) it also follows that
if we are given an infinite number of copies and asked to delete a single
copy, then we can accomplish it without any error. This is in accordance with
our fundamental understanding about quantum information. For $2$-to-$1$
deleting the optimal quality is $0.70$.

{\it Conditional quantum deleting of qubits:}
To study how the quantum information is distributed among various subsystems
and imperfections introduced during a deletion process, we introduce a
special class of deleting machines the so-called conditional deleting.
If the two input qubits are identical then machine deletes a copy and if
they are different then it allows them to pass through
without any change.
For orthogonal qubits we define it as
$ \arrowvert 0 \rangle\arrowvert 0 \rangle \arrowvert A \rangle \rightarrow
\arrowvert 0 \rangle\arrowvert \Sigma \rangle \arrowvert A_0 \rangle,~
\arrowvert 1 \rangle\arrowvert 1 \rangle \arrowvert A \rangle \rightarrow
\arrowvert 1 \rangle\arrowvert \Sigma \rangle \arrowvert A_1 \rangle,~
\arrowvert 0 \rangle\arrowvert 1 \rangle \arrowvert A \rangle \rightarrow
\arrowvert 0 \rangle\arrowvert 1 \rangle \arrowvert A \rangle,$~
and $\arrowvert 1 \rangle\arrowvert 0 \rangle \arrowvert A \rangle
\rightarrow \arrowvert 1 \rangle\arrowvert 0 \rangle \arrowvert A \rangle$,
where $\arrowvert A \rangle$ is the initial state and $\arrowvert A_0 \rangle,
\arrowvert A_1 \rangle$ are the final states of ancilla. Notice that if the
conditional deletion for orthogonal qubits has 
to work it is necessary to include the ancilla. Then for an arbitrary qubit
the deleting operation will create the following state

\begin{eqnarray}
&& \arrowvert \Psi \rangle \arrowvert \Psi \rangle \arrowvert A \rangle = [\alpha^2 \arrowvert 00 \rangle
+ \beta^2 \arrowvert 1 1 \rangle+
\alpha \beta ( \arrowvert  0 1 \rangle+ \arrowvert  1 0 \rangle) ] \arrowvert  A \rangle\nonumber \\
&& \rightarrow \alpha^2 \arrowvert  0 \rangle \arrowvert \Sigma \rangle \arrowvert  A_0 \rangle
+ \beta^2 \arrowvert 1 \rangle \arrowvert \Sigma \rangle \arrowvert  A_1 \rangle+
\alpha \beta ( \arrowvert  0 1 \rangle+ \arrowvert  1 0 \rangle)  \arrowvert A \rangle \nonumber \\
&& = \arrowvert\Psi_{{\rm out}} \rangle.
\end{eqnarray}
However, ideally for an arbitrary qubit the deleting machine should have
created $\arrowvert \Psi \rangle\arrowvert\Psi \rangle\arrowvert A
\rangle \rightarrow \arrowvert \Psi \rangle \arrowvert\Sigma \rangle
\arrowvert A_{\Psi} \rangle$.
Since the final output states in ideal and actual cases are different
linearity does not allow to conditionally delete an unknown quantum state.


{\it State-dependent and approximate deleting machine:}
Since it is impossible to delete an unknown state perfectly we may ask how
well one can do the above operation. Here, we discuss the approximate
deleting of an unknown state and the fidelity of a state-dependent
quantum deleting machine. In (13) if we wish to delete a qubit then the
ancilla state should belong to a three dimensional Hilbert space.
For a normalised output state in (13) (and hence unitary)
we need $\arrowvert A \rangle, \arrowvert A_0 \rangle,$ and $\arrowvert
A_1 \rangle$ to be orthogonal to each other.
The reduced density matrix of the two qubits $ab$ after the deleting operation
is given by

\begin{eqnarray}
&& \rho_{ab} = {\rm tr}_{c} (\arrowvert\Psi_{\rm out} \rangle\langle\Psi_{\rm out}  \arrowvert) =
\arrowvert\alpha \arrowvert^4 \arrowvert 0 \rangle \langle 0  \arrowvert \otimes \arrowvert\Sigma \rangle\langle\Sigma \arrowvert\nonumber \\
 && + \arrowvert\beta \arrowvert^4 \arrowvert 1 \rangle\langle1  \arrowvert \otimes
  \arrowvert\Sigma \rangle\langle\Sigma \arrowvert
 + 2 \arrowvert\alpha \arrowvert^2  \arrowvert\beta \arrowvert^2  \arrowvert\psi^+ \rangle\langle\psi^+  \arrowvert,
\end{eqnarray}
where $\arrowvert \psi^{+} \rangle= \frac{1}{\sqrt 2} (\arrowvert 01 \rangle+ \arrowvert 10 \rangle)$ is
one of the four maximally entangled states.
The reduced density matrix for the qubit in the mode $b$ will be

\begin{eqnarray}
 \rho_{b} = {\rm tr}_{a} (\rho_{ab}) =
(1 - 2 \arrowvert\alpha \arrowvert^2  \arrowvert\beta \arrowvert^2 ) \arrowvert\Sigma \rangle\langle\Sigma \arrowvert
 +  \arrowvert\alpha \arrowvert^2  \arrowvert\beta \arrowvert^2  I,
\end{eqnarray}
where $I$ is the identity matrix in two dimensional Hilbert space.
Thus the reduced density matrix of the qubit in the mode $b$ is a mixed
state which contains the error due to imperfect deleting. The fidelity
of deleting can be defined
as $F_b =   \langle\Sigma \arrowvert\rho_{b} \arrowvert\Sigma \rangle= 
(1 -  \arrowvert\alpha \arrowvert^2  \arrowvert\beta \arrowvert^2 )$.
This shows that for either $\alpha = 0$ and $\beta = 1$ or  $\alpha = 1$ and
$\beta = 0$ the fidelity of deleting is maximum.
For an equal superposition of qubit state the fidelity
reaches $\frac{3}{4}$ which is the maximum limit for deleting an
unknown qubit. The average fidelity of deleting is given by ${\bar F}_b =
\int d \Omega  F_b = \frac{5}{6} \approx 0.83$, where $\Omega= \sin \theta d\theta d \phi$.

  We can see how good the state of the qubit in mode $a$ is after both 
the qubits have passed through a quantum deleting machine. The reduced
density matrix of this mode is given by

\begin{eqnarray}
&& \rho_{a} = {\rm tr}_{b} (\rho_{ab}) =
 \arrowvert\alpha \arrowvert^4  \arrowvert 0 \rangle\langle  0 \arrowvert
 +  \arrowvert\beta \arrowvert^4  \arrowvert 1 \rangle\langle  1 \arrowvert + \arrowvert\alpha \arrowvert^2
 \arrowvert\beta \arrowvert^2 I. \nonumber \\
\end{eqnarray}
The fidelity of the qubit in mode $a$ is $ F_a =   \langle\Psi \arrowvert\rho_{a}
\arrowvert\Psi \rangle= (1 -  2\arrowvert\alpha \arrowvert^2  \arrowvert\beta \arrowvert^2 )$. 
For an equal superposition of qubit state the
fidelity is $\frac{1}{2}$.
The average fidelity in this case is ${\bar F}_a = \int d \Omega
F_a = \frac{2}{3} \approx 0.66$.
This shows that the first mode of the qubit is not faithfully retained
during the deleting operation. It is, in fact, less than the actual deleting
mode. This shows that  linearity of quantum theory neither allows us
to delete an unknown state perfectly nor does retain the
original state of the other qubit. We can compare the quantum deleting operation
to that of the quantum cloning operation defined by Wootters and Zurek
\cite{wz}. In the cloning operation the reduced density matrix of both the
modes are same \cite{bh}.
Therefore, the average fidelities of both the modes are found to be
$\frac{2}{3}$.
However, as we have shown here the fidelity of the
two modes are different for the deleting operation.
This again suggests us that the quantum deleting machine is not 
the reverse of quantum cloning machine.

{\it Quantum deleting of non-orthogonal states:}
In some physical situations two qubits need not be in orthogonal states nor
in completely arbitrary states but they could be chosen secretly from a set
containing non-orthogonal states.
Though each of a copy from two copies of two orthogonal states
can be perfectly deleted, we show here that the same cannot be done for
two non-orthogonal states. Suppose we have two copies of the
two of the non-orthogonal states $\arrowvert\Psi_i \rangle, (i =1,2)$ with a finite
scalar product between them. We ask if there is a unitary
operator which can delete one of the copy by keeping the other intact.
For simplicity and clarity we work without attaching an ancilla to the qubits.
For two copies of distinct non-orthogonal states the deleting machine is
a unitary operator which acts on the combined Hilbert space of two qubits
and would create the following transformation:
$\arrowvert\Psi_1 \rangle\arrowvert\Psi_1 \rangle \rightarrow \arrowvert\Psi_1 \rangle\arrowvert\Sigma \rangle,~
\arrowvert\Psi_2 \rangle\arrowvert\Psi_2 \rangle \rightarrow \arrowvert\Psi_2 \rangle\arrowvert\Sigma \rangle,~
\arrowvert\Psi_1 \rangle\arrowvert\Psi_2 \rangle \rightarrow \arrowvert\Psi_1 \rangle\arrowvert\Psi_2 \rangle,~$
and $\arrowvert\Psi_2 \rangle\arrowvert\Psi_1 \rangle \rightarrow \arrowvert\Psi_2 \rangle\arrowvert\Psi_1 \rangle$.
Since unitary evolution must preserve the {\it inter inner products}
we have several conditions to be satisfied  simultaneously. 
These restriction are $\langle\Psi_1 \arrowvert\Psi_2 \rangle^2 =  \langle\Psi_1 \arrowvert\Psi_2 \rangle$,
$\langle \Psi_1 \arrowvert \Psi_2 \rangle = \langle \Sigma \arrowvert \Psi_2 \rangle$,
$\langle\Sigma \arrowvert \Psi_2 \rangle=1$, $\langle\Sigma \arrowvert \Psi_1 \rangle= 1$, and
$\langle\Psi_2 \arrowvert \Psi_1 \rangle= \langle\Sigma \arrowvert \Psi_1 \rangle$. These can be
satisfied only if $\arrowvert\Psi_1  \rangle= \arrowvert \Psi_2 \rangle= \arrowvert\Sigma \rangle$, which
means a contradiction as there are no non-trivial states being processed by
the machine. Thus, copies of non-orthogonal states cannot be deleted by a
unitary machine.

{\it Perfect deleting and signalling:}
We know that if we could clone an arbitrary state we can send superluminal
signals \cite{dd,ng,akp1}.
The natural question is whether no-deleting principle is also consistent with
no-signalling. At prima face, it looks that it may not be so. But on the
other hand linear structure of quantum mechanics and no-signalling
are consistent with each other and no-deleting is a consequence of linearity.
So there should be some nontrivial link between no-deleting and no-signalling.
Below we show how this works.

Suppose Alice and Bob share two pairs of EPR singlets and Alice has particles
$1$ and $3$ and Bob has $2$ and $4$. Since the singlet state is invariant
under local unitary operation $U \otimes U$ it is same in all basis.
Let us write the combined state of the system in an arbitrary qubit basis
$\{ \arrowvert\psi \rangle = \cos \theta \arrowvert 0 \rangle +
\sin \theta \arrowvert 1 \rangle, \arrowvert{\bar \psi} \rangle =
\sin \theta \arrowvert 0 \rangle - \cos \theta \arrowvert 1 \rangle \}$
                              
\begin{eqnarray}
\arrowvert\Psi^{-} \rangle_{12} \arrowvert\Psi^{-} \rangle_{34} =
\frac{1}{2}(\arrowvert\psi \rangle_1 \arrowvert\psi \rangle_3
\arrowvert{\bar \psi} \rangle_2 \arrowvert{\bar \psi} \rangle_4 +
\arrowvert{\bar \psi} \rangle_1 \arrowvert{\bar \psi} \rangle_3 \arrowvert\psi \rangle_2 \arrowvert\psi \rangle_4 \nonumber \\
- \arrowvert{\bar \psi} \rangle_1 \arrowvert\psi \rangle_3 \arrowvert\psi \rangle_2 \arrowvert{\bar \psi} \rangle_4
- \arrowvert\psi \rangle_1 \arrowvert{\bar \psi} \rangle_3 \arrowvert{\bar \psi} \rangle_2 \arrowvert\psi \rangle_4 ).
\end{eqnarray}
Now if Alice measures her particles $1$ and $3$ onto the basis
$\arrowvert\psi \rangle_1 \arrowvert\psi \rangle_3$, then the Bob's
particles $2$ and $4$ are in the state $\arrowvert{\bar \psi} \rangle_2
\arrowvert{\bar \psi} \rangle_4 $. If Alice measures her particles in the
basis $\arrowvert{\bar \psi} \rangle_1 \arrowvert{\bar \psi} \rangle_3$, then
Bob's particles are in the state $\arrowvert\psi \rangle_2 \arrowvert\psi
\rangle_4$. Similarly, one can find the resulting states with other choices of
measurements. So whatever measurements Alice does, if she does not convey
the measurement result to Bob, then Bob's particles are in a completely
random mixture (i.e. $\rho_{24} = \frac{I_2}{2} \otimes \frac{I_2}{2}$ ).
But suppose Bob has a conditional quantum deleting machine, which can delete
an arbitrary state. Then, after Alice does measurement he attaches an ancilla
and deletes his copies. The four possible choices for the states of the four
particles (with ancilla) are give by

\begin{eqnarray}
&& \arrowvert{\bar \psi} \rangle_1 \arrowvert{\bar \psi} \rangle_3  \arrowvert\psi \rangle_2 \arrowvert\psi \rangle_4 \arrowvert A \rangle
 \rightarrow \arrowvert {\bar \psi} \rangle_1 \arrowvert {\bar \psi} \rangle_3  \arrowvert \psi \rangle_2 \arrowvert\Sigma \rangle_4  \arrowvert A_{\psi} \rangle\nonumber \\
&& \arrowvert\psi \rangle_1 \arrowvert\psi \rangle_3 \arrowvert{\bar \psi} \rangle_2 \arrowvert{\bar \psi} \rangle_4 \arrowvert A \rangle
 \rightarrow \arrowvert\psi \rangle_1 \arrowvert\psi \rangle_3 \arrowvert{\bar \psi} \rangle_2 \arrowvert\Sigma \rangle_4  \arrowvert A_{{\bar \psi}} \rangle \nonumber \\
&& \arrowvert{\bar \psi} \rangle_1 \arrowvert\psi \rangle_3 \arrowvert\psi \rangle_2 \arrowvert{\bar \psi} \rangle_4 \arrowvert A \rangle
 \rightarrow \arrowvert{\bar \psi} \rangle_1 \arrowvert\psi \rangle_3 \arrowvert\psi \rangle_2 \arrowvert{\bar \psi} \rangle_4 \arrowvert A \rangle  \nonumber \\
&& \arrowvert\psi \rangle_1 \arrowvert{\bar \psi} \rangle_3 \arrowvert{\bar \psi} \rangle_2 \arrowvert\psi \rangle_4 \arrowvert A \rangle
 \rightarrow \arrowvert\psi \rangle_1 \arrowvert{\bar \psi} \rangle_3 \arrowvert{\bar \psi} \rangle_2 \arrowvert\psi \rangle_4 \arrowvert A \rangle.
\end{eqnarray}
If we trace out ancilla and particles $1$ and $3$ the reduced density
matrix for Bob's particles $2$ and $4$ is given by

\begin{eqnarray}
&& \rho_{24} = \frac{1}{4} \bigg(
\arrowvert \psi \rangle_2{_2}\langle \psi \arrowvert \otimes
\arrowvert \Sigma \rangle_4{_4}\langle \Sigma \arrowvert+
\arrowvert{\bar \psi} \rangle_2{_2}\langle{\bar \psi}
\arrowvert \otimes \arrowvert \Sigma \rangle_4{_4}\langle \Sigma \arrowvert+
\nonumber \\
&& \arrowvert\psi \rangle_2{_2}\langle \psi \arrowvert \otimes
\arrowvert {\bar \psi} \rangle_4{_4}\langle {\bar \psi} \arrowvert+
\arrowvert{\bar \psi} \rangle_2{_2}\langle{\bar \psi} \arrowvert \otimes
\arrowvert\psi \rangle_4{_4}\langle\psi \arrowvert\bigg)
\end{eqnarray}
which clearly depends on the choice of basis.
This shows that if Alice measures her particles in $\{ \arrowvert0 \rangle, \arrowvert1 \rangle\}$
basis then Bob's particles are in one density matrix. If Alice measures her
particles in $\{ \arrowvert+ \rangle, \arrowvert- \rangle\}$ basis Bob's particles are left in
a different density matrix. Therefore, if Bob can delete an arbitrary state
he can distinguish two statistical mixtures  and will allow superluminal
signalling. Hence, Bob cannot delete an arbitrary state.
Therefore, quantum no-deleting principle is in unison with the principle of
no-signalling.

Quantum no-deletion principle being a fundamental limitation on quantum
information it ought to have some implications.
For example, this may provide special security to copies of files
in a quantum computer and possibly in quantum cryptographic protocols--
which deserve further investigation in the future. However, constructing a
universal quantum deleting machine and obtaining the optimal fidelity of
quantum deletion is still an open problem.

Financial support from EPSRC is gratefully acknowledged.
We thank useful discussions at various stages with C. Bennett, S. Bose,
N. Cerf, A. Chefles and N. Gisin.

\vskip 1cm

\renewcommand{\baselinestretch}{1}


\end{document}